\documentclass{iaus}
\usepackage{graphics}

  \checkfont{eurm10}
  \iffontfound
    \IfFileExists{upmath.sty}
      {\typeout{^^JFound AMS Euler Roman fonts on the system,
                   using the 'upmath' package.^^J}%
       \usepackage{upmath}}
      {\typeout{^^JFound AMS Euler Roman fonts on the system, but you
                   dont seem to have the}%
       \typeout{'upmath' package installed. iaus.cls can take advantage
                 of these fonts,^^Jif you use 'upmath' package.^^J}%
      }
  \else
  \fi

  \checkfont{msam10}
  \iffontfound
    \IfFileExists{amssymb.sty}
      {\typeout{^^JFound AMS Symbol fonts on the system, using the
                'amssymb' package.^^J}%
       \usepackage{amssymb}%
         \let\leq=\leqslant
         
      }{}
  \fi

  \IfFileExists{amsbsy.sty}
    {\typeout{^^JFound the 'amsbsy' package on the system, using it.^^J}%
     \usepackage{amsbsy}}
    {}

\newsavebox{\astrutbox}
\sbox{\astrutbox}{\rule[-5pt]{0pt}{20pt}}

\newcommand\etal{\mbox{\textit{et al.}}}

\newcommand{\gta}{\gtrsim}
\renewcommand{\deg}{^{\circ}}

\title[A dichotomy in dust-jet orientation in radio galaxies]{A dichotomy in dust-jet orientation in radio galaxies}

\author[Gijs Verdoes Kleijn and Tim de Zeeuw]%
{Gijs Verdoes Kleijn$^1$%
\and Tim de Zeeuw$^2$}

\affiliation{$^1$ESO, Karl-Schwarzschild-strasse 2, Garching, 85748,Germany, email: gverdoes@eso.org\\ [\affilskip]
$^2$Leiden Observatory, Postbus 9513, NL-2300 RA Leiden, Netherlands.}

\pubyear{2004}
\volume{222}
\pagerange{1--8}
\date{?? and in revised form ??}
\jname{The Interplay among Black Holes, Stars and ISM \\in Galactic Nuclei}
\editors{Th. Storchi Bergmann, L.C. Ho \& H.R. Schmitt, eds.}
\begin{document}

\maketitle

\begin{abstract}

We have analyzed the position angle (PA) differences between radio jets and dust distributions in the centers of Fanaroff \& Riley Type
1 (FRI) radio galaxies. We model the observed PA differences to infer
the three-dimensional relative orientation of jet and dust. Our main
conclusion is that there is a dichotomy in dust-jet-galaxy orientation both in
projection and in three-dimensional space. 
The orientation dichotomy can
explain the contradictory results obtained in previous studies. We
briefly mention scenarios that might explain the dichotomy.

\end{abstract}


The relative orientations of the jets, dust and
stellar host pose constraints on the physical processes that govern the
jet orientation and perhaps the jet formation mechanism in radio
galaxies. Many studies have found that jets in radio galaxies are
roughly perpendicular to the dust structures (e.g., Kotanyi \& Ekers
1979; M\"{o}llenhoff, Hummel \& Bender 1992; van Dokkum \& Franx 1995;
Verdoes Kleijn \etal\ 1999; de Koff \etal\ 2000; de Ruiter \etal\ 2002; Capetti \& Celotti 1999;
Sparks \etal\ 2000). In contrast, Schmitt \etal\ (2002) found for a
sample of 20 radio galaxies with regular dust disks that the jets are
not at all perpendicular to the disks in {\sl three-dimensional}
space.

We have compiled a sample of 37 FR-I dusty radio galaxies at $z<0.15$ with published radio-jet orientations for which we could find HST archival imaging to determine the intrinsic relative orientation of (ir)regular dust, jet and stellar host. 


We classify the dust morphology in three bins of increasing
irregularity. There are 17 'ellipses' which have a smooth elliptical circumference with ellipticity $\epsilon$, 9 'lanes', which are thin filamentary structures with bends and/or mulitple strands. Nine galaxies have a rather regular dust morphology intermediate between ellipse and lane. Lastly, there are 2 'irregular  dust' features  which are too clumpy and irregular to establish an orientation.  
We define the dust orientation as the PA of the longest linear
axis for lanes 
or of the dust major axis for ellipses. 

Figure 1a shows a dichotomy in the PA difference between dust and galaxy major axis
$\Delta$PA$_{\rm DG}$ as a function of the PA difference between dust
and jet axis $\Delta$PA$_{\rm DJ}$. 
Dust structures which are misaligned with the galaxy ($\Delta$PA$_{\rm DG} >
20\deg$) are typically 'lanes' and have roughly perpendicular jets ($\Delta$PA$_{\rm DJ} \gta 60\deg$).
In contrast, the radio jets have no clear preferential direction (i.e., $\Delta$PA$_{\rm DJ}\sim [25\deg-90\deg]$) to dust structures which are aligned with the galaxy ($\Delta$PA$_{\rm DG} < 20\deg$) which are typically classified as 'ellipses'. 



The sky-plane projected properties of the dust-jet system, $\Delta$PA$_{\rm DJ}$ and $\epsilon$, place constraints on the three-dimensional relative orientations. The dust ellipses are consistent with circular thin disks
observed at random viewing angles. We therefore model these system as circular disks for which the angle $0\deg \leq \theta_{\rm DJ} \leq 90\deg$ between jet-axis and the disk rotation-axis follows a Gaussian distribution. This model gives the simultaneous probability distribution for ($\Delta$PA$_{\rm DJ}$,$\epsilon$) for a given mean $\mu$ and dispersion $\sigma$. Figure~1b shows the best-fitting $\mu$ and $\sigma$ for 100 bootstrapping runs using a maximum-likelihood analysis. In contrast to ellipses, we do not know
the inclination of dust lanes. The systematic difference in $\Delta$PA$_{\rm DG}$ and morphology argue against the lanes being simply edge-on ellipses. Dust lanes could be perturbed disks, viewed relatively close to edge-on, as has been argued for two dust lanes in the sample, NGC 5128 and M84 (Quillen \etal\ 1992; Quillen \& Bower 1999). This interpretation requires that face-on lanes are classified differently, i.e., as 'irregular dust' and/or round 'ellipses'. Thus we model lanes as edge-on systems, also because this will yield an upper-limit on the width of their $\theta_{\rm DJ}$ distribution. Figure~1b shows that the dust lanes have 
systematically smaller $\theta_{\rm DJ}$ than dust ellipses. 

At least two scenarios could explain qualitatively the orientation dichotomy. First, the dust is acquired externally, initially forms a perturbed disk (observed as a dust lane) and finally settles into a flat disk (i.e., a dust ellipse). This scenario would imply that jets form roughly aligned with the initial angular momentum of the $\sim$kpc-scale dust. 
Second, it might be that radio-jet induced pressure gradients in the ambient gas force some disks to be roughly perpendicular to the jets (and hence generally misaligned with the galaxy potential), creating disk warps and perturbations in the process (Quillen \& Bower 1999). A full analysis of the radio-jet orientation dichotomy and its cause(s) is in progress.  

\begin{figure}
\centering
\resizebox{6cm}{!}{\includegraphics{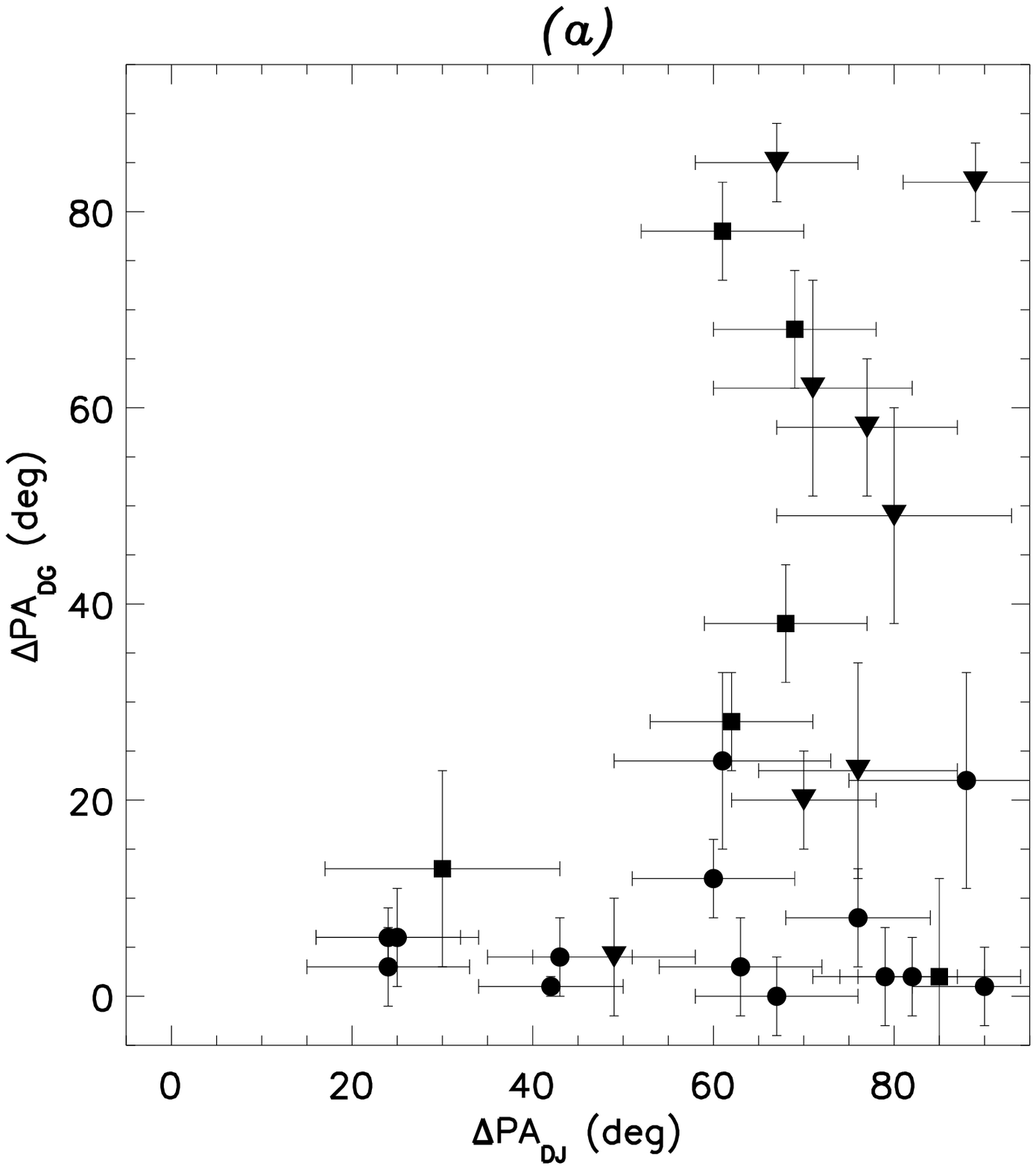} }
\resizebox{6cm}{!}{\includegraphics{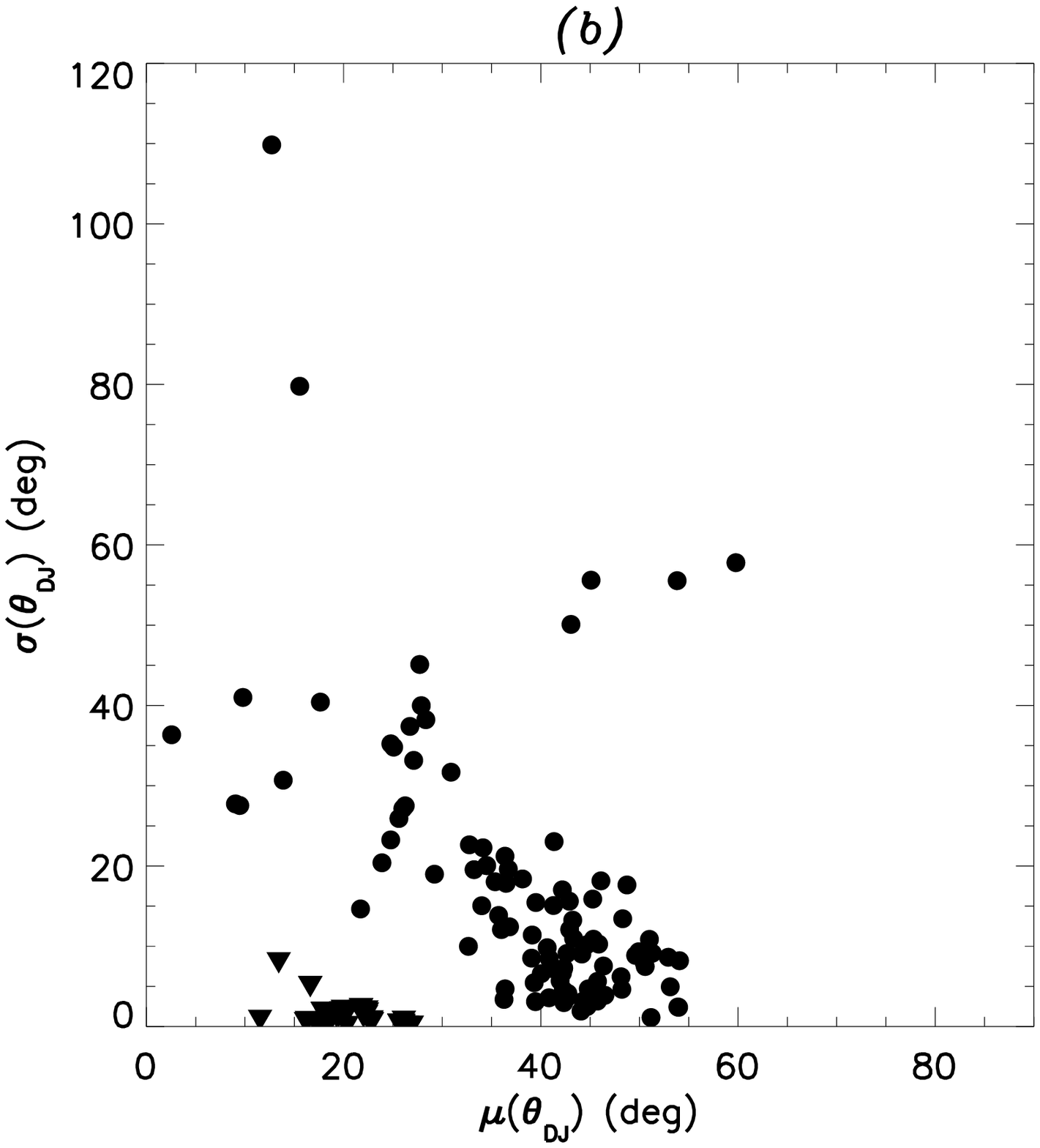} }
\caption[]{{\rm (a)}:projected PA differences for Dust, Galaxy and Jet. $\Delta$PA$_{\rm DG}$ versus $\Delta$PA$_{\rm DJ}$ for dust ellipses (dots), lanes (triangles) and morphologies between ellipse and lane (squares). (b): 3D misalignment angle $\theta_{\rm DJ}$ between jet-axis and disk rotation-axis. Best fitting mean $\mu$ versus dispersion $\sigma$ for a truncated Gaussian distribution of $\theta_{\rm DJ}$ for a 100 bootstrapping runs using a maximum-likelihood analysis for ellipses (dots) and lanes (triangles). Lanes have a narrower $\theta_{\rm DJ}$ distribution than dust ellipses.}
\end{figure}

\end{document}